\documentclass{egpubl}
\usepackage{eg2023}

\ConferencePaper        
\usepackage[T1]{fontenc}
\usepackage{dfadobe}  
\usepackage{amsmath}
\DeclareMathOperator*{\argmin}{arg\,min}
\usepackage{cite}  
\BibtexOrBiblatex
\electronicVersion
\PrintedOrElectronic
\ifpdf \usepackage[pdftex]{graphicx} \pdfcompresslevel=9
\else \usepackage[dvips]{graphicx} \fi

\usepackage{xcolor}
\usepackage{siunitx}
\usepackage{egweblnk} 
\usepackage{gensymb}

\usepackage{amsfonts}

\usepackage[a-2u]{pdfx}

\DeclareMathOperator*{\render}{\mathfrak{R}}

\newcommand{\cTI}[1]{}
\newcommand{\cTR}[1]{}
\newcommand{\cTN}[1]{}
\newcommand{\cAW}[1]{}
\newcommand{\cES}[1]{}
\newcommand{\cMM}[1]{}

\title[Automatic inference of anatomically meaningful solid wood texture from a single photograph]%
      {Automatic inference of a anatomically meaningful solid wood texture from a single photograph}

\author[Thomas K. Nindel \& Mohcen Hafidi \& Tomáš Iser, Alexander Wilkie]
{\parbox{\textwidth}{\centering Thomas K. Nindel\thanks{Corresponding author}$^{1,2}$\orcid{0000-0002-8003-2672}, Mohcen Hafidi$^{1}$\orcid{0000-0002-1931-0493}, Tomáš Iser$^{1}$\orcid{0000-0003-3013-8994},
        and Alexander Wilkie$^{1}$\orcid{0000-0002-3536-6577} 
        \newline
        \newline
        \centering{$^1$ Charles University, Faculty of Mathematics and Physics, Czech Republic}
        \newline
        \centering{$^2$Berufsakademie Sachsen, Dresden, Germany}
        }
}

\volume{42}   

\begin{document}

 \teaser{
  \includegraphics[trim={0 0 0 0},clip,width=1.0\linewidth]{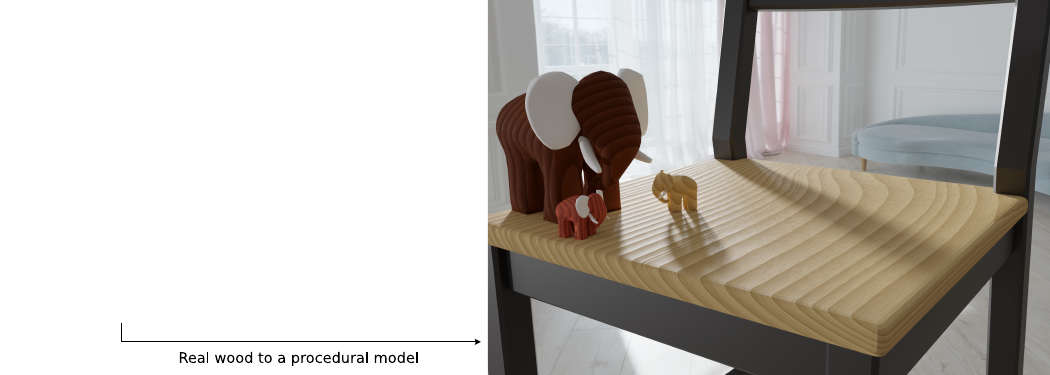} 
  \centering
   \caption{Our method is capable of taking a 2D~scan of a real wood specimen (left, photograph) and then automatically matching corresponding parameters for a procedural solid wood appearance model based on \cite{liu2016}. Such a model is fully three-dimensional and allows realistic rendering of carved wooden solids (right, Monte Carlo simulated render) based on the input photographs.
   }
 \label{fig:teaser}
}

\maketitle
\begin{abstract}
Wood is a volumetric material with a very large appearance gamut that is further enlarged by numerous finishing techniques. Computer graphics has made considerable progress in creating sophisticated and flexible appearance models that allow convincing renderings of wooden materials. 
However, these do not yet allow fully automatic appearance matching to a concrete exemplar piece of wood, and have to be fine-tuned by hand. More general appearance matching strategies are incapable of reconstructing anatomically meaningful volumetric information. This is essential for  applications where the internal structure of wood is significant, such as non-planar furniture parts machined from a solid block of wood, translucent appearance of thin wooden layers, or in the field of dendrochronology.

In this paper, we provide the two key ingredients for automatic matching of a procedural wood appearance model to exemplar photographs: a good initialization, built on detecting and modelling the ring structure, and a phase-based loss function that allows to accurately recover growth ring deformations and gives anatomically meaningful results.
Our ring-detection technique is based on curved Gabor filters, and robustly works for a considerable range of wood types.


\end{abstract}

\section{Introduction}

\begin{figure}[b]
  \centering
  \includegraphics[clip,width=\linewidth]{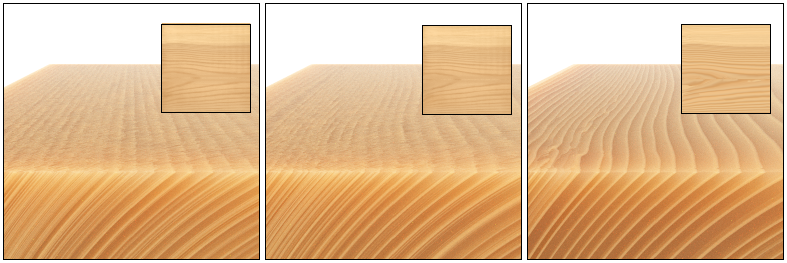}
  \caption{False-color volume rendering of a solid texture that was fit using a differentiable material graph, (left and middle) compared to our approach. The density of the volume is set to $\sigma_t=15$ for growth-lateness of greater $0.9$, $\sigma_t=0.5$ otherwise, both with a homogeneous single-scattering albedo $\alpha=(0.805, 0.578, 0.359)^T$ (RGB). The model in the first image was optimized with our board orientation estimate, but without an intial guess for the deformation field as a prior. The second image additionally uses our initial deformation field. The third image shows a visualization of our result; insets are the respective frontal, non-volume renderings. Notice how in spite of the very good match in terms of frontal appearance, the left and middle volumes show blurry ring boundaries.}
  \label{fig:overfit}
\end{figure}

In \emph{wood rendering} and \emph{wood appearance modeling}, the current state of the art is based either on highly precise BSSRDF measurements (BTFs, appearance scans), or procedural models.
Measurement-based approaches yield pixel-perfect matches at the cost of high storage and memory-bandwidth requirements, acquisition complexity, and inability to edit the data after acquisition.
On the other hand, procedural wood models allow artistic control and editing, but they are difficult to match to given wood samples, both if an exact match to a given piece of wood is needed, or only with regard to a general wood type.
Recent work on using optimisation to match procedural material models to observations \cite{shi2020:tog} has been quite successful in a broad range of settings, including wood.
However, as figure~\ref{fig:overfit} shows, current approaches still fail to properly match a given wood sample \emph{down to its internal 3D structure} -- a feature that is needed to make wood grain wrap correctly around a solid 3D object. Specifically, as figure~\ref{fig:overfit} shows, extant techniques are capable of generating an internal ring structure - but not one that really convincingly matches the wood grain pattern seen on the top surface, like in the results we show for our technique in figure~\ref{fig:results}.

Our main contribution, described in section~\ref{sect:method}, is that we propose a robust, deterministic method for the automatic inference of a locally fit, procedural 3D material model for wood. Our approach allows one to obtain realistic solid wood textures that can be carved to any geometry (such as the elephants in Fig.~\ref{fig:teaser}). Our approach is not based on machine learning and does not rely on any training datasets.
%

\begin{figure}[tb]
  \centering
  \includegraphics[width=\linewidth]{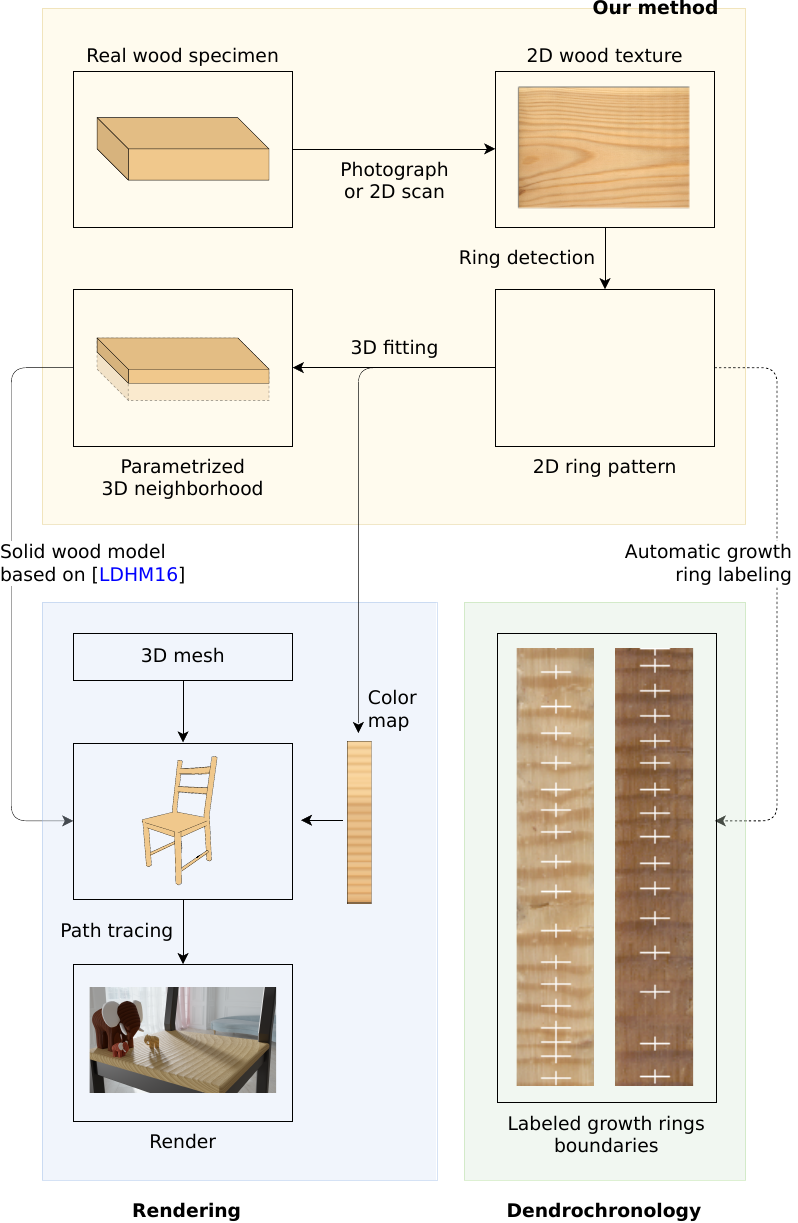}
  \caption{Method pipeline overview. Our method starts with a real wood specimen (top left) that is scanned or photographed from one side. The 2D~image is then processed to detect the wood growth rings (see also Fig.~\ref{fig:pipeline}). This process is already useful for automatic ring labeling in dendrochronology (bottom right), where the input image is a 2D~scan of a flattened drilled cylinder of a tree. For use in solid wood rendering (bottom left), the 2D~ring pattern has to be further processed and fitted in a 3D~space, since we are interested in parameters of a solid 3D~appearance model based on \cite{liu2016}. As only a 2D~scan is available as input, the 3D~fitting can only be expected to approximate the actual specimen in its local neighborhood. The final path-traced render is using the fitted parameters and model on an arbitrary user-specified 3D~mesh.}
   \label{fig:method_diagram}
\end{figure}

Our work builds on the assumption that by accurately identifying the basic ring structure and its deformations, we can satisfy not only the requirements from wood rendering, that is to attain closely matching structural appearance (Fig.~\ref{fig:method_diagram}, bottom left), but also the requirements of \emph{dendrochronology} for a precisely aligned identification of ring boundaries (Fig.~\ref{fig:method_diagram}, bottom right).
In the dendrochronology field (section~\ref{sect:dendro}), core samples of living trees are inspected for their growth rings variations, and the resulting data is interesting for climate research, archaeology, and art history. Automatic approaches have to be robust to the wide range of wood anatomical features, as certain wood species (e.g., diffuse-porous hardwoods) are very difficult to work with. Our approach performs well even in these settings, and in fact, we used it in dendrochronological contexts as a means of verifying its performance.

\section{Related work}
%
\subsection{Modelling wood appearance}
%

Wood is a complex material with an intricate volumetric structure. Hartmann et al. \cite{hartmann2017modelling} dive into the complex process that controls the formation of wood cells, whose variance is ultimately responsible for the emergence of the characteristic growth patterns. From a technical standpoint, wood is a bio-composite formed from three different kinds of molecules: cellulose, hemicellulose, and lignin. Lignin is the main chemical responsible for the wood color. It is preferentially found in areas of thick cell walls \cite{li2018optically}, which explains why late growth shows a higher color saturation than early growth in coniferous species. Chemically replacing lignin with a transparent polymer results in transparent wood, a material with interesting optical properties \cite{vasileva2018light}. 

The basic approaches for modelling wood in photorealistic rendering can be divided into \emph{material scanning} ones, which essentially treat the material as a surface of certain optical properties, and \emph{procedural}, which typically model the whole three-dimensional wood interior.
Our method is based on finding the corresponding parameters for a procedural model, such that a scanned surface of a wood can be transformed into a fully procedural and editable 3D~model.

\paragraph*{Material scanning and BTFs}
The appearance of a specific specimen of wood can be measured and encapsulated in various ways. Bidirectional Texture Functions (BTFs) \cite{dana1999reflectance} store the full 6-dimensional surface reflectance using, essentially, lookup tables. Material scanners are available that capture diffuse reflectance, spatially varying roughness and a normal map, which can be fed into an appropriate BRDF implementation, such as the Disney Principled BRDF \cite{burley2012physically}. Both approaches can yield impressive realism for the specimen they were applied to, but suffer from immense storage requirements.
 Henzler et al. \cite{henzler2020learning} propose an interesting data-driven approach to generate solid textures from 2D specimen, wood being one of their use-cases. Their approach has the advantage of providing an infinite solid texture, but represents the input texture only qualitatively.
\paragraph*{General solid noise functions}
Procedural noise can be a versatile tool to capture the texture of a wide range of materials, Perlin~Noise \cite{perlin1985image} and Gabor~Noise (e.g. \cite{lagae2009procedural}) being prominent examples. Matching the noise function parameters to obtain a desired input texture is also possible: For example, \cite{galerne2012gabor} show a method to match the power spectrum of Gabor~Noise to exemplars, while \cite{BangaruMichel2021DiscontinuousAutodiff} show an example of fitting the parameters of a Perlin~Noise even in the presence of discontinuities. Since our goal is to closely match the ring structure, using power-spectrum based noise would require additional and hard to control steps, since matching the rings requires a precise match of both the noise signals' amplitude and \emph{phase}.

\paragraph*{Domain-specific procedural models}
A state of the art appearance model of wood is the work of Liu et al. \cite{liu2016}, on whose work we based our results (Fig.~\ref{fig:method_diagram}). Their model uses various distortion functions to modulate a cylindrical coordinate system. Together with a sawtooth-like function that describes the color saturation changes in the earlywood / latewood transitions, they also provide expressions for diffuse reflectance and the fibre direction. These can be used as base functions to define inputs to a BRDF shader. By the explicit way they model wood “age” on a sub-ring scale, their model can be seen as anatomically informed. Once the model parameters are tuned, a process that requires manual work by an expert, it is very expressive and can be used to model many species of wood. 
We are directly building on this model, and our contribution is that we are able to automatically recover its main parameters. We also recover a color map that describes coarse-level variations as a function of ring growth time.

Larsson et al. \cite{larsson2022procedural} extend the expressive power of wood-specific procedural solid textures by the integration of wood knots, based on a skeletal description of the branches and a distance-field based formulation of their respective area of influence to the growth rings formation.
%
\subsection{Procedural appearance matching}
\label{sec:appearance-matching}
%
Differentiable material graphs allow for the use of gradient-based optimization to polish material parameters, after a good initial guess has been found ~\cite{guo2020bayesian, shi2020:tog}. Since the success of appearance matching depends on (and can be limited by) the expressive power of the underlying procedural model, neural loss functions \cite{aittala2016reflectance} seem to be best suited in finding plausible parameters for a wide range of material graphs, because they evaluate appearance differences in qualitative, perceptual terms. This allows these algorithms to converge to satisfactory results even if the expressiveness of the underlying graph is limited. However, none of these works provide a 3D solid texture that can be used to "carve" objects from or accurately replicate the curvature of the growth rings present in the target.

\cite{lefebvre2000analysis} show a half-automatic system that can extract parameters for a procedural 3D solid wood texture. From a thresholded input image, they are able to estimate ring frequency and spatial orientation of a given wood plank, based on image statistics. Ring variations are modelled statistically in terms of turbulence intensity and frequency. Their orientation estimates are based on statistical properties of the thresholded image, while ours is based on accurately identified rings. Our method also accurately recovers the deformation field.
%
%
\subsection{Dendrochronology}
%
\label{sect:dendro}
Automated dendrochronological measurements rely on image processing techniques that use either photos \cite{wang2010automatic,fabijanska2016comparative,fabijanska2017towards,lenty2020tree}, or computed-tomography (CT) data \cite{martinez2021automated} as their input. Salient points of these methods is the application of 2D image processing techniques
to the problem of growth ring identification. They achieve reasonable accuracy for coniferous species, but their accuracy greatly degrades in the presence of pores. 

Notable data driven approaches include \cite{FABIJANSKA2018353}, who use a U-Net that was trained on manually labeled data, while \cite{Polavek2022} use a Mask~R-CNN architecture. Neural networks seem to improve detection rates compared to the approaches based on traditional 2D image processing especially for ring porous wood, a category that our ring detection method shows on-par performance. Data driven approaches create likelihood maps for ring boundaries, that are subsequently thresholded to get exact locations. In contrast, our approach directly pinpoints the ring boundaries without the ambiguity of a likelihood.

Martinez et al. \cite{martinez2021automated} extend the idea of tree ring extraction to 3D data and reconstruct tree ring isosurfaces from X-ray computed tomography data based on edge detection. Basing the method on CT data makes the approach very robust, because of strong correlation between X-ray intensities and material density. The limited availability of CT scanners hinders the practical applicability of their approach.

In a more broader context, systems for the identification of tree species from either microscopic or macroscopic images build on segmenting anatomical features of wood, including its rings. Martins \cite{martins2018towards} contains a survey of the proposed methods.
Datasets of dendrochronological measurements are available through the work of \cite{fabijanska2017towards}, the \textit{WIAD} database \cite{rademacher2021wood}, both containing labels. The DendroElevator \cite{DendroElevator} database accumulates data from several sources, containing some very detailed tree ring images (up to microscopic scale) obtained from wood core samples. Some of their data are labelled.
%
\subsection{Finding rings in images}
%
Another field that considers the enhancement, identification, and segmentation of ring-like structures is \emph{fingerprint identification}. The key differences to our domain are the bandwidth of ring frequencies (fingerprint ridges are quite evenly spaced), and the presence of branching (fingerprint ridges can split and merge, which usually is not the case for tree rings). This bandwidth limitation makes transferring techniques that are informed by or based on frequency-domain information (e.g., \cite{thai2016filter, le2020fingerprint}) somewhat more problematic, but not impossible. To our knowledge, the connection between the fingerprint enhancement domain and growth ring identification was first discovered in \cite{jonsson2008detection}.
We base our approach on \emph{curved Gabor filters} \cite{gottschlich2011curved}, which perform very well on fingerprint images. Gabor filters \cite{gabor1946theory} have been used across many disciplines for signal analysis and processing, and are often amongst the top performers from the suitable candidates. A very interesting property is their capability for retrieving phase information \cite{kong2009analysis}, further \cite{grohs2019stable, alaifari2021gabor}.

\begin{figure*}[htb]
  \centering
  \includegraphics[trim={0.27cm 0 0.27cm 0},clip,width=\linewidth]{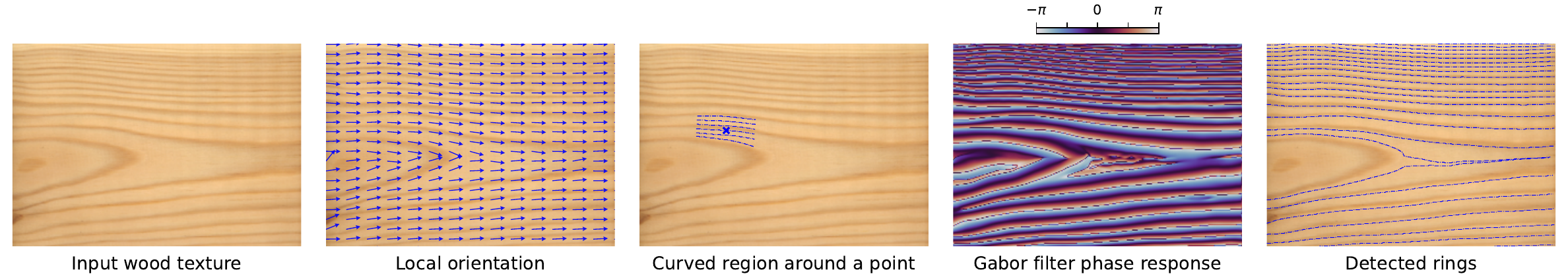}
  \caption{Ring detection process. We start with an input 2D~texture (left) and compute its local orientation. We then compute a curved region around each point in the image and apply a Gabor filter on it. Accumulating over all pixels gives the Gabor filter phase response. We then trace the rings in the phase image to detect their positions (right).}
  \label{fig:pipeline}
\end{figure*}

\section{Method}
\label{sect:method}
\subsection{Overview}

Our overall goal (Figs.~\ref{fig:teaser},~\ref{fig:method_diagram}) is to reconstruct a subset of parameters that we can plug into the procedural wood appearance model proposed by Liu et al. \cite{liu2016}. Their starting point is a cylindrical coordinate system that has the z-axis aligned with the stem of the tree, and they apply distortion functions that work on both the radial distance and the azimuth. They further propose several versions of the spatially varying distortion field, like procedural noise, or using a lookup table (``distortion texture''). This can be either applied in a radially symmetric way, where the texture-space y-coordinates correspond to radial distances and the x-coordinates to values of z in tree space, or using a more elaborate helical wrapping to allow distortions that vary with the azimuth.

Our approach aims at recovering this distortion texture, and a basic color map, from a single photograph of the tangential plane of a wood cut. Please also refer to Figure~\ref{fig:pipeline} for a graphical overview.

We recover this data using a three-stage process. First, we calculate both an  estimate of the board orientation, and an initial estimate of the distortion field from the locations of growth rings (Sec.~\ref{sec:ringdetect}, \ref{sec:loc}). The growth rings are found using curved Gabor filters (Sec.~\ref{sec:cgf}). Second, we polish the distortion field using a signal-phase informed optimization scheme (Sec.~\ref{sec:polish}). Finally, a color map is calculated.

%
%

\subsection{Curved Gabor filters}
\label{sec:cgf}
The key idea to curved Gabor filters is to calculate them in an image over curved regions that follow local orientation (Fig.~\ref{fig:pipeline}, left to middle). The initial local orientation is obtained from the 2D image gradient: we use the Scharr operator \cite{scharr2000optimal} after smoothing the image with a small Gaussian kernel. The gradient is averaged over a rectangular~neighborhood, rotated $90\si{\degree}$ and normalized
to obtain the initial local orientation field.

Once the local orientation is known, we compute a curved region around each point in the image.
Each curved region consists of $2p+1$ parallel contours with $2q+1$ points each, in our implementation $p=q=80$.
They are constructed by first sampling \emph{perpendicular} to the local orientation for $p$~steps, into both positive and negative directions. From each of the $2p+1$ points, a sampling walk \emph{along} the local orientation in positive and negative directions for $q$~steps, gives the pixel values of the curved region.
As the step size is 1 pixel for both, we can look at the result as a pixelated patch of $2p \times 2q$ pixels that follows the local curvature.

Because the patch is rotated so that it follows the local orientation, growth rings will always be aligned horizontally when sampled isometrically. This allows estimating the local ring-frequency of a patch from the number of peaks is contains. 

Using the (fixed) ring orientation inside each patch~$P$ and the detected frequency, we then calculate the convolution of the patch with a complex Gabor kernel $K=\{K_\mathrm{re}, K_\mathrm{im}\}$ of an appropriate size to get the filtered patch $P’$:

\begin{equation}
P' = P \ast K,
\end{equation}
\begin{align}
K_\mathrm{re}(x,y,\theta, f, \sigma_x, \sigma_y) &=
  \exp\bigg(
    - \frac{1}{2}
    \bigg[\frac{x_\theta^2}{\sigma_x^2}
  +
  \frac{y_\theta^2}{\sigma_y^2}\bigg]
\bigg)
\cdot
\cos(2\pi f x_\theta),
\\
K_\mathrm{im}(x,y,\theta, f, \sigma_x, \sigma_y) &=
  \exp\bigg(
    - \frac{1}{2}
    \bigg[\frac{x_\theta^2}{\sigma_x^2}
  +
  \frac{y_\theta^2}{\sigma_y^2}\bigg]
\bigg)
\cdot
\sin(2\pi f x_\theta),
\end{align}
\begin{equation}
\begin{aligned}
x_\theta &= x \cdot \cos \theta + y \cdot \sin \theta,
\\
y_\theta &= -x \cdot \sin \theta + y \cdot \cos \theta.
\end{aligned}
\end{equation}

We accumulate the resulting signals $P'$ over all patches, giving a complex valued filtered image $I_f=(I_\mathrm{re}, I_\mathrm{im})$. Transforming to polar coordinates gives a magnitude-image $I_\mathrm{mag}=\sqrt{I_\mathrm{re}^2 + I_\mathrm{im}^2}$, and a phase-image $I_\phi=\arctan_2({I_\mathrm{im}, I_\mathrm{re}})$. From the phase information, we can estimate the location of tree ring boundaries at where $I_\phi(x, y)=\frac{\pi}{2}$, the location of the signals negative zero crossing. The phase image contains a sign ambiguity that we resolve later (Sections~\ref{sec:polish},~\ref{sec:dendro-results}).

As as side-effect to calculating the Gabor filter on curved regions, it is also possible to estimate local curvature and local ring frequency. The local curvature can be computed from the absolute value of the differences of the local orientation of the contours' middle points and their end points. This can be used to identify anomalies such as knots.

The final local orientation field is obtained from the phase image using the same procedure that calculated the initial local orientation estimate, which gives a smooth field that follows along the growth ring's trajectory.
%
\subsection{Ring detection}
\label{sec:ringdetect}
%
%
To detect a growth ring, we seed the algorithm at any point $p=(x,y)$ where $I_\phi(x,y)\approx{}\frac{\pi}{2}$, and walk along the local orientation field into both positive and negative directions. By following the orientation field instead of just thresholding the phase image, our algorithm can trace rings even across small anomalies.
%
\subsection{Estimating board location}
\label{sec:loc}
With the detected visible growth rings it becomes possible to reason about the boards orientation within the 3D~tree coordinate system. Using this positional information then allows us to calculate an estimate of the distortion field that 
would lead to the observed pattern. We show this for tangential cut planes, but the approach can easily be extended to other cuts as well.

Our position estimate starts with automatically finding the projection of the tree center (the $z$~axis) onto the plane. It is usually located between the two rings with the largest mean distance from one another, with the remaining inter-ring distances being monotonically decreasing to both sides. This projection of the tree-center also gives the translation of the tangential board along its y-axis. From the median average inter-ring distance, we can induce the average scale of the rings in image-space coordinates. To find the x-translation of the observed plane, we fit the radial scaling factor and translation on the XY-plane using a brute force search in the neighbourhood of the initial estimate, which only takes seconds to compute. The resulting sign of the x-translation is ambiguous since we are only observing a plane.
\subsection{Modeling growth distortions and initial guess}
\label{sec:radial}

In their model, Liu et. al. \cite{liu2016} model distortions of points in tree space $q$ in the radial and/or tangential direction, magnitudes $m_r$ and $m_t$. The magnitudes are spatially varying and can be any function $f\colon R^3 \to R$, such as a Perlin noise, or a texture. The location of the distorted point $q'$ is given by

\begin{equation}
    q' = d(p) = q + m_r(q)\hat{r} + m_t(q)\hat{t}.
\end{equation}

$\hat{r}$ and $\hat{t}$ are unit vectors into the radial and tangential direction, respectively. As the initial guess for the distortion field, we use the difference between the observed and the ideal, undistorted rings in image space and map this displacement to radial distortions $m_r(q)\hat(r)$ using the board location estimate. Tangential distortions $m_t(q)$ are not used here due to the ambiguity of radial and tangential distortions when only observing a planar projection of the deformed tree rings.

\subsection{Polishing the initial deformation field}
\label{sec:polish}

A sufficiently accurate initial guess of the deformation field is important for accurate convergence of the final polishing step. The periodic nature of the texture function constrains gradient-based optimization to the interval of the period the starting point is contained in. More precisely, the initial deformation field needs to have a local phase error of less than $\frac{\pi}{2}$ for best results.

Formally, we want to minimize an energy $E$ that is a function of the reference image $J$ and a rendered image $I$.

\begin{equation}
    \argmin_{m_r, c, s_r} E(J, I)
\end{equation}

$I$ is obtained from a rendering operator, $\render$, that takes the boards orientation, given by the linear transformation $T$, a ring scaling factor $s_r$, the distortion field $d(q)$ and a color map $c(q_r)$ to an rendered image $I$:

\begin{equation}
    I = \render(T, s_r, d, c)
\end{equation}

$s_r$ encodes the mean width of growth rings in radial units, the color map $c$ describes, for each ring individually, the color-changes observable as the tree growth forms early and late growth. The distortion field encodes the ring growth variations. In its simplest form, the rendering function is just a projection of the solid texture to image space by a cutting plane. Note that all functions need to be differentiable in order for gradient descent optimization to be applicable.


%

\paragraph*{Loss function}
Optimizing using a per-pixel loss, or a feature loss, will result in a good appearance matched \emph{image}, even good estimates for BRDF parameters other than color, through the use of a differentiable renderer as $\render$. Problems arise due to the co-optimization of both distortion field and color map and will lead to a fit that lacks anatomical meaning (Fig.~\ref{fig:overfit}).  This is due to the non-orthogonality of the parameters: From a reasonable gamut of colors in the color map, individual surface pixels can be modulated by using the distortion field, making it a proxy to achieve a certain surface color. 
This does not move the rings as a whole, but rather individual points. The result is a non-smooth, "fuzzy" deformation field that can even lead to ring fold-over.

%
To overcome this, the optimization needs to be constrained, or the metric improved. We have experimented implementing a monotonic constraint on the distorted radii to discourage foldover, and we also tried enforcing smoothness by convolving the distortion field with a Gaussian kernel in exponentially increasing intervals during optimization. However, we were not able to attain significant improvements.
Another potential issue arises from the periodic discontinuity the procedural model contains, which is a direct result of the underlying sawtooth-like function that describes the growth periods of non-tropical wood grain (cf. figure~\ref{fig:dendro-plot}). It has been shown that not accounting for discontinuities can lead to convergence issues (shown for example by Loubet et. al. in \cite{Loubet2019Reparameterizing}) . We implemented an experimental reparameterizing renderer on top of the Mitsuba~3 renderer \cite{jakob2022mitsuba3} to correctly handle texture discontinuities, and also tried smoothing the falling edge of the yearly growth boundary by convolving with a small Gaussian kernel. Neither lead to satisfactory convergence behaviour in our experiments.

We finally opted for improving the optimization metric by introducing using a loss based on signal phase. We also separate the optimization of the deformation field from the color map into a two stage process. Our loss encodes the difference in signal phase between points on the reference image, and points on the rendered image of the current iteration.

\begin{equation}
    E = \big(J_{\phi} - I_{\phi}\big)^2
\end{equation}


The phase-based loss function lets the optimizer fit the locations of entire rings holistically. One could implement a phase-based loss on top of a bank of regular Gabor filters that recover the phase information for both the reference image $I$ and the rendering of the current iteration. Propagating loss gradients back through this filter bank is very costly, however. Due to the constrained nature of the underlying solid texture (tree rings cannot be arbitrarily placed, but follow certain rules), the phase information can be directly obtained: rendering the fractional part of the radial coordinate in the distorted tree's cylindrical coordinate system, multiplied by $2\pi$ circumvents this bottleneck. The phase response of the curved Gabor filter calculated from the input image earlier then serves as the optimization target. 

The phase ambiguity on the reference image can be resolved by assuming a fixed orientation, and flipping the phase values by 180° beyond the projected tree center, given it is visible in the image. We also note that the loss function of the optimization needs to respect the cyclic nature of phase information when taking differences.
\paragraph*{Optimization algorithm and hyperparameters}
We implemented the fitting procedure using a custom differentiable rasterizer. The rasterizer is built on top of TensorFlow \cite{tensorflow2015-whitepaper}, which is used as a framework for reverse-mode automatic differentiation. We use their implementation of the Adam optimizer \cite{kingma2014adam} to drive the optimization loop. The learning rate was set to $0.03$, and we run the optimization for $500$ epochs. A single run takes less than a minute on a NVIDIA RTX 2080 GPU for a target texture with a resolution of $750 x 735$.

\begin{figure}[b]
  \centering
  \includegraphics[trim={0.0 3.2cm 0.0cm 2.0cm},clip,width=\linewidth]{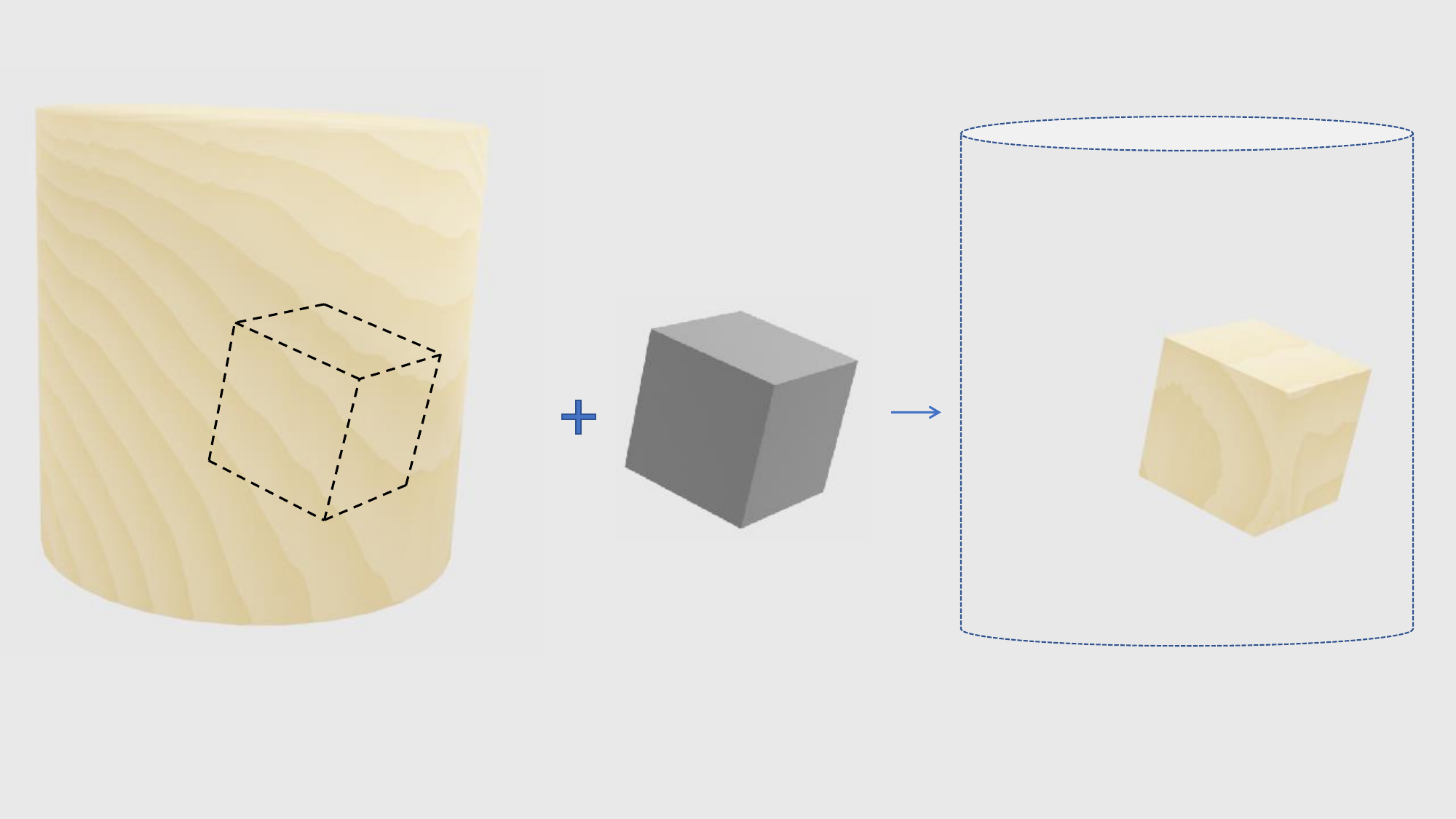}
  \caption{The procedural wood model is an infinite 3D~texture centered around a virtual tree (left). An arbitrary 3D~mesh, such as a cube (middle), can be positioned within the 3D~texture coordinate system accordingly to where the object was carved from the tree. This gives a textured solid wood object (right) that can be rendered.}
  \label{fig:woodenCube}
\end{figure}

\begin{figure}[b]
  \centering
  \includegraphics[clip,width=\linewidth]{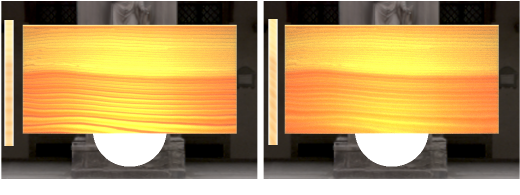}
  \caption{Volumetric appearance is highly dependent on grain orientation - shown here are renderings of 1.8mm thick veneer sheets backlit by a spherical light source. The volumetric properties are derived from our fit (right), and extruded along the surface normal (left). Insets show side-views to illustrate grain}
  \label{fig:vol-compare}
\end{figure}

\begin{figure}[tb]
  \centering
  \includegraphics[clip,width=\linewidth]{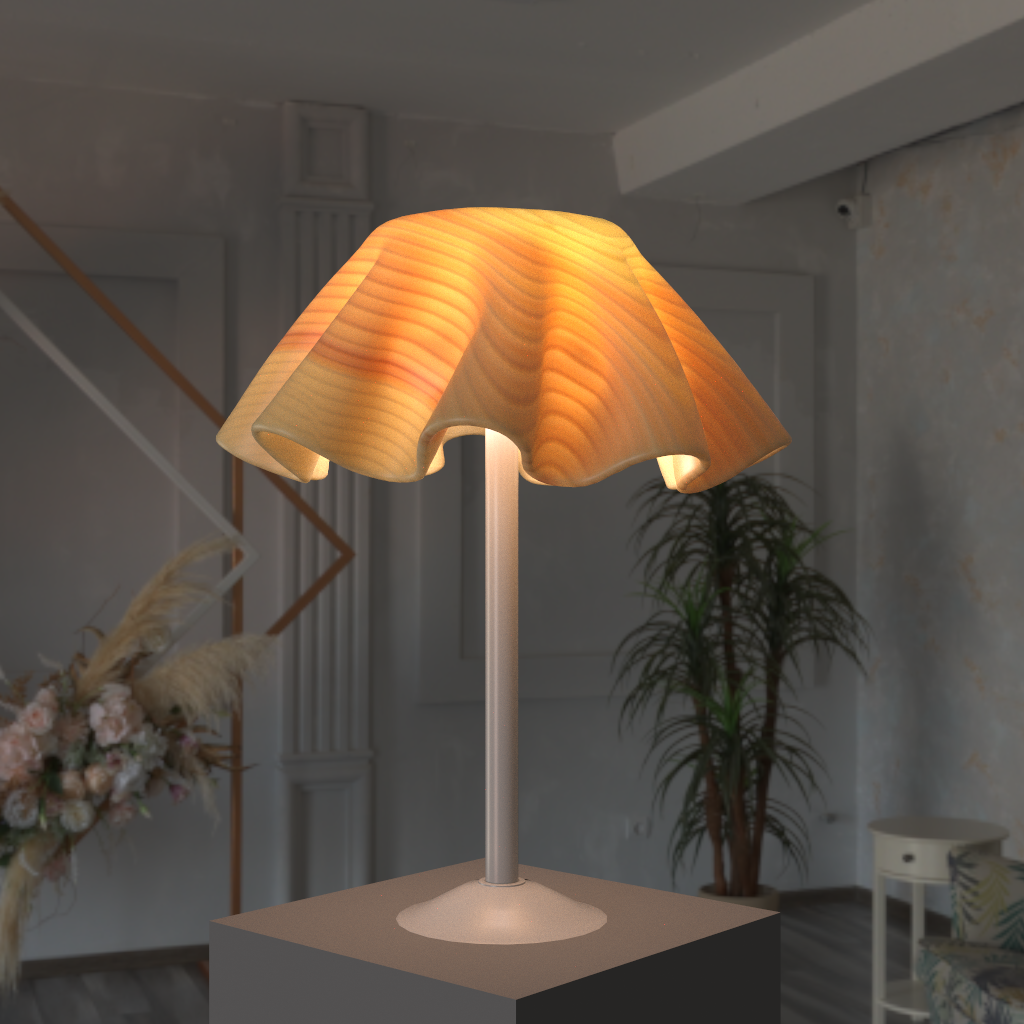}
  \caption{Volumetric rendering of a lampshade that was cut from a fit model using our method (Fig.~\ref{fig:results}, 2nd column). Both single scattering albedo and volumetric density are heterogeneous.}
  \label{fig:vol-lampshade}
\end{figure}

\begin{figure*}[htb]
  \centering
  \includegraphics[clip, width=\linewidth, trim={0.4cm 0 0 0}]{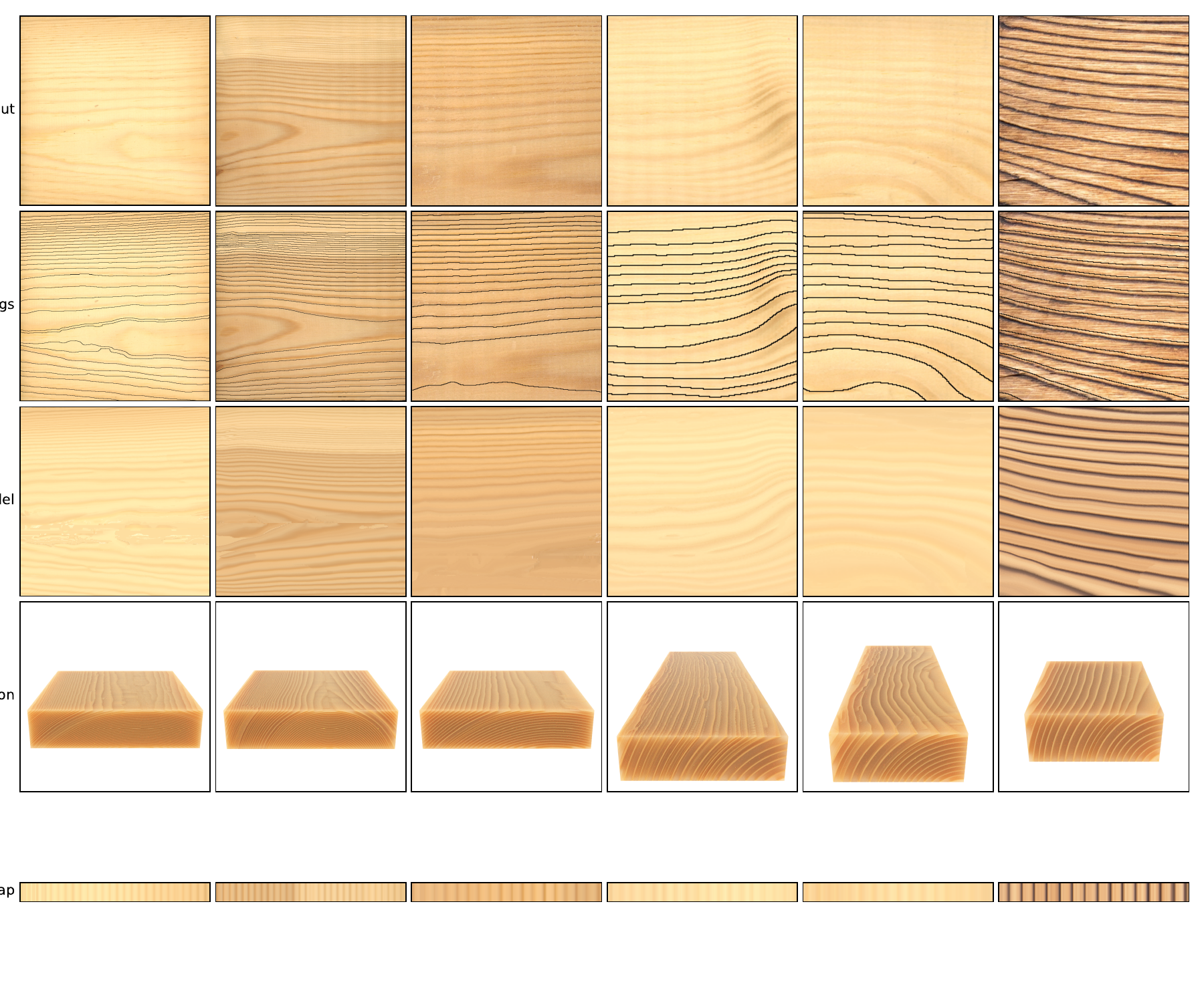}
  \caption{Results obtained with our approach. Rows from top to bottom: Input texture, detected rings, frontal view of fit model (diffuse reflectance only), volumetric visualization (false color), fit color map.}
  \label{fig:results}
\end{figure*}
\section{Results}

\subsection{Structurally matching wood appearance}

We apply our method to several pieces of coniferous wood, where most rings are identified correctly. Difficulties arise where the plane grazes a growth ring tangentially, where tiny variations of the estimated board location and the ring distortion can create huge changes in appearance. This can be seen in Figure~\ref{fig:results}, 1st and 3rd column, where mistakenly identified rings add to the problem, leading to a noisy, implausible distortion estimate in that area. In contrast, the specimen in the third column does not suffer from this problem, since it does not contain the projection of the tree center.

Through the fit, we obtain a solid texture (see also Figure~\ref{fig:woodenCube}) that also supplies \emph{semantic information} that we use to derive other BRDF parameters. In Figure~\ref{fig:teaser}, we show a rendering that modulates the roughness and displacement as a function to the lateness of growth of each ring, which results in a very realistic appearance with specular highlights that are in alignment with the late growth.

\subsection{Volumetric appearance}

For small material thicknesses, volumetric light transport becomes significant for the appearance of wood. Examples are decorative use-cases such as veneer lamp shades, or functional usages such as translucent wood touch panels, or architectural use of delignified wood. 
To illustrate the importance of subsurface reconstruction of the wood grain, we simplify the cellular anatomy into an approximate participating medium. Using our fit, we set the spatially varying density proportionally to the lateness of growth, latewood being more dense than earlywood. The volumetric properties were estimated to a scattering coefficient of $\mu_t\approx16.0 mm^{-1}$ for latewood, $\mu_t\approx6.5 mm^{-1}$ for earlywood, and a (wavelength dependent) single scattering albedo derived from the surface reflectance. 

To come up with the density parameters, we applied the Inverse~Adding-Doubling~Method \cite{prahl2011everything} on data provided in \cite{sugimoto2018reflection}, who performed measurements on Sugi wood, a coniferous subspecies, using micro spectrometer hardware. 
The heterogeneous single scattering albedo is calculated using the surface albedo mapping function proposed by Elek et. al \cite{ElekSumin2017SGA}, their Eqn.~4. 
Fig.~\ref{fig:vol-compare} compares the translucent appearance of both a model that was fit using our pipeline, and a simple extrusion of the surface ring pattern into the depth of the volume. The appearance difference is explained by the orientation of the latewood shells, which is perpendicular in the extruded case, but follows wood grain in the fit case. In Fig.~\ref{fig:vol-lampshade} we show a volumetric rendering of a lampshade. The characteristic warm-colored translucent appearance wood shows when it is backlit is captured very well.

\subsection{Dendrochonology}
\label{sec:dendro-results}
We confirm the robustness of our ring detection method by running it on a dendrochonological dataset. There, accuracy of detection is very important, since it is used as primary data for various other fields, including climate research.

\begin{figure}[h]
  \centering
  \includegraphics[clip,width=\linewidth]{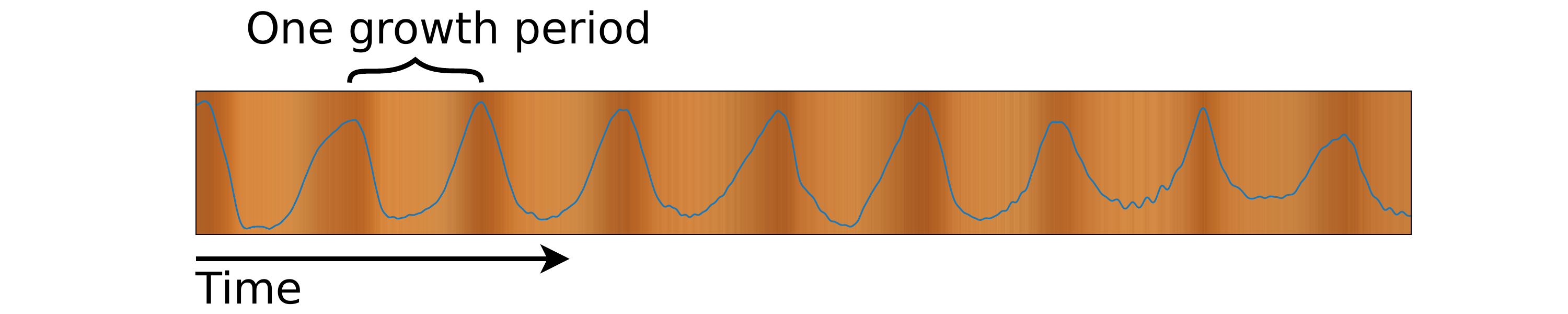}
  \caption{From the extracted color map, it becomes possible to gain insight into the specifics of each growth cycle, here the latewood-transition. The plot shows the normalized and inverted lightness value of the HSL-transformed colorbar, smoothed by a gaussian kernel. The lightness is proportional to the cell-wall thickness and lignin content.}
  \label{fig:dendro-plot}
\end{figure}

To detect rings in dendrochonolgy images, the images are first being scaled to uniform width, and then converted to $HSV$ colorspace, from which only the V-channel is used. The detection procedure is executed as detailed above. The phase sign ambiguity is resolved by assuming an image orientation with the youngest rings at the bottom of the image. The recovered rings are then compared to ground-truth by measuring the closest distance between the ground-truth label (one x/y coordinate pair for each ring). If the distance is smaller than $3$ pixels, conforming with the evaluation shown in \cite{fabijanska2017towards}, the ring is counted as a match. Table~\ref{tab:dendro} summarizes the performance of our approach. Performance for coniferous woods is good. For ring-porous woods, an anatomical variety that has shown to be problematic in previous work, our results are exellent. The high accuracy of our fit means that it can also supply additional data for dendrochronological evaluations. Fig.~\ref{fig:dendro-plot} shows a partial color map, superimposed with the inverse lightness, which correlates with the time the latewood transition occurred in the respective year.

%
%
%

\begin{table}[]
\begin{tabular}{ll|ll|ll|ll}
        &       & \textbf{Ours} &      & \multicolumn{2}{l|}{\cite{FABIJANSKA2018353}} & \multicolumn{2}{l}{\cite{fabijanska2017towards}} \\
species & rings & sen  & pre & sen        & pre & sen                   & pre \\ \hline
larch   & 99    & 0.77 & 0.83 &            &      & 0.90                  & 1.00 \\
spruce  & 115   & 0.93 & 0.96 &            &      & 0.99                  & 1.00 \\
pine    & 108   & 0.79 & 0.91 &            &      & 0.93                  & 0.97 \\
ash     & 114   & 0.98 & 0.97 & 0.97       & 0.98 & 0.45                  & 0.85 \\
birch   & 30    & 0.17 & 0.06 &            &      & 0.86                  & 0.74
\end{tabular}
\caption{Comparison of sensitivity (sen) and precision (pre) of our ring-detection method and two state of the art methods on a dendrochronological dataset. \cite{FABIJANSKA2018353} was developed especially with ringporous wood in mind. Our method performs especially well on ringporous wood (ash) and gives decent results on confierous species (first three rows). Diffuseporous woods such as birch are problematic, where our approach produces many false positives.}
\label{tab:dendro}
\end{table}



\section{Conclusion}
%
In this paper, we demonstrated curved Gabor filters to be an efficient technique for the extraction of tree ring information from single images of planar wooden boards. This information can then be used to build a realistic, anatomically meaningful 3D~procedural texture that closely matches the structure of the log the original board was cut from.

The main focus of our work was to establish an automatic pipeline that is fast and accurate. The technology we ported and adapted from fingerprint detection not only serves as a tree ring detector with great performance, but also enables the efficient optimization of the ring deformations based on signal phase. The reconstruction of additional BSDF parameters by using differentiable rendering could further improve the model: but as the results of this paper show, the approach we currently use is already suitable for production work. Our pipeline further enables volumetric rendering of wood's subsurface light transport, and can pave a way to finding a faster, approximate BSSRDF, which we see as a interesting direction for further work.

We also see a potential of our robust ring extraction technique to be adopted by the dendrochronology community, where ring labelling is an important task that our method solves as a side effect.

\section{Acknowledgements}
This work has received funding from GA~UK project 148222 of Charles University. This work was further supported by the Charles University grant SVV-260699, and from the European Union’s
Horizon 2020 research and innovation program under the Marie
Sklodowska-Curie grant agreement No. 956585.
%
%

\bibliographystyle{eg-alpha-doi} 
\bibliography{wood-paper}       


\newpage

\end{document}